\documentclass[aps,preprint,superscriptaddress,longbibliography]{revtex4-1}

\usepackage{graphicx}
\usepackage{dcolumn}
\usepackage{bm}
\usepackage{amssymb, amsmath}
\usepackage{microtype}
\usepackage{xfrac}
\usepackage{array}
\usepackage{gensymb}
\usepackage{xcolor}
\usepackage[normalem]{ulem}
\usepackage{dsfont}
\usepackage{diagbox}

\usepackage[T1]{fontenc}

\usepackage{indentfirst} 
\usepackage{float}
\usepackage{graphicx}
\usepackage{ulem}
\usepackage{setspace}
\usepackage{physics}

\newcommand{\kbf}{\ensuremath{\bm k}}

\newcommand{\ef}{\textrm{$E_F$}}

\newcommand{\kz}{$k_z$}

\newcommand{\rtrt}{$\sqrt{3}\times \sqrt{3}$}

\begin{document}
 
\title{Reversible Non-Volatile Electronic Switching in a Near Room Temperature van der Waals Ferromagnet}

\author{Han Wu}
\affiliation{Department of Physics and Astronomy and Rice Center for Quantum Materials, Rice University, Houston, TX, 77005 USA}

\author{Lei Chen}
\affiliation{Department of Physics and Astronomy and Rice Center for Quantum Materials, Rice University, Houston, TX, 77005 USA}

\author{Paul Malinowski}
\affiliation{Department of Physics, University of Washington, Seattle, Washington 98195, USA}

\author{Jianwei Huang}
\affiliation{Department of Physics and Astronomy and Rice Center for Quantum Materials, Rice University, Houston, TX, 77005 USA}

\author{Qinwen Deng}
\affiliation{Department of Physics and Astronomy, University of Pennsylvania, Philadelphia, PA 19104, USA}

\author{Kirsty Scott}
\affiliation{\mbox{Department of Physics, Yale University, New Haven, Connecticut 06520, USA}}
\affiliation{\mbox{Energy Sciences Institute, Yale University, West Haven, Connecticut 06516, USA}}

\author{Bo Gyu Jang}
\affiliation{Theoretical Division and Center for Integrated Nanotechnologies, Los Alamos National Laboratory, Los Alamos, NM, USA}

\author{Jacob P. C. Ruff}
\affiliation{Cornell High Energy Synchrotron Source, Cornell University, Ithaca, NY 14853, USA}

\author{Yu He}
\affiliation{Department of Applied Physics, Yale University, New Haven, Connecticut 06511, USA}

\author{Xiang Chen}
\affiliation{Department of Physics, University of California, Berkeley, Berkeley, California 94720, USA}

\author{Chaowei Hu}
\affiliation{Department of Physics, University of Washington, Seattle, Washington 98195, USA}
\affiliation{Department of Materials Science and Engineering, University of Washington, Seattle,
Washington 98195, USA}

\author{Ziqin Yue}
\affiliation{Department of Physics and Astronomy and Rice Center for Quantum Materials, Rice University, Houston, TX, 77005 USA}

\author{Ji Seop Oh}
\affiliation{Department of Physics, University of California, Berkeley, Berkeley, California 94720, USA}
\affiliation{Department of Physics and Astronomy and Rice Center for Quantum Materials, Rice University, Houston, TX, 77005 USA}

\author{Xiaokun Teng}
\affiliation{Department of Physics and Astronomy and Rice Center for Quantum Materials, Rice University, Houston, TX, 77005 USA}

\author{Yucheng Guo}
\affiliation{Department of Physics and Astronomy and Rice Center for Quantum Materials, Rice University, Houston, TX, 77005 USA}

\author{Mason Klemm}
\affiliation{Department of Physics and Astronomy and Rice Center for Quantum Materials, Rice University, Houston, TX, 77005 USA}

\author{Chuqiao Shi}
\affiliation{Department of Materials Science and NanoEngineering, Rice University, Houston, TX, 77005, USA}

\author{Yue Shi}
\affiliation{Department of Physics, University of Washington, Seattle, Washington 98195, USA}

\author{Chandan Setty}
\affiliation{Department of Physics and Astronomy and Rice Center for Quantum Materials, Rice University, Houston, TX, 77005 USA}

\author{Tyler Werner}
\affiliation{Department of Applied Physics, Yale University, New Haven, Connecticut 06511, USA}

\author{Makoto Hashimoto}
\affiliation{Stanford Synchrotron Radiation Lightsource, SLAC National Accelerator Laboratory, Menlo Park, California 94025, USA}

\author{Donghui Lu}
\affiliation{Stanford Synchrotron Radiation Lightsource, SLAC National Accelerator Laboratory, Menlo Park, California 94025, USA}

\author{T. Yilmaz}
\affiliation{National Synchrotron Light Source II, Brookhaven National Lab, Upton, New York 11973, USA}

\author{Elio Vescovo}
\affiliation{National Synchrotron Light Source II, Brookhaven National Lab, Upton, New York 11973, USA}

\author{Sung-Kwan Mo}
\affiliation{Advanced Light Source, Lawrence Berkeley National Laboratory, Berkeley, CA 94720, USA}

\author{Alexei Fedorov}
\affiliation{Advanced Light Source, Lawrence Berkeley National Laboratory, Berkeley, CA 94720, USA}

\author{Jonathan Denlinger}
\affiliation{Advanced Light Source, Lawrence Berkeley National Laboratory, Berkeley, CA 94720, USA}

\author{Yaofeng Xie}
\affiliation{Department of Physics and Astronomy and Rice Center for Quantum Materials, Rice University, Houston, TX, 77005 USA}

\author{Bin Gao}
\affiliation{Department of Physics and Astronomy and Rice Center for Quantum Materials, Rice University, Houston, TX, 77005 USA}

\author{Junichiro Kono}
\affiliation{Department of Materials Science and NanoEngineering, Rice University, Houston, TX, 77005, USA}
\affiliation{Departments of Electrical and Computer Engineering, Rice University, Houston, TX, 77005, USA}
\affiliation{Department of Physics and Astronomy and Rice Center for Quantum Materials, Rice University, Houston, TX, 77005 USA}

\author{Pengcheng Dai}
\affiliation{Department of Physics and Astronomy and Rice Center for Quantum Materials, Rice University, Houston, TX, 77005 USA}

\author{Yimo Han}
\affiliation{Department of Materials Science and NanoEngineering, Rice University, Houston, TX, 77005, USA}

\author{Xiaodong Xu}
\affiliation{Department of Physics, University of Washington, Seattle, Washington 98195, USA}
\affiliation{Department of Materials Science and Engineering, University of Washington, Seattle,
Washington 98195, USA}

\author{Robert J. Birgeneau}
\affiliation{Department of Physics, University of California, Berkeley, Berkeley, California 94720, USA}
\affiliation{Materials Sciences Division, Lawrence Berkeley National Laboratory, Berkeley, California 94720, USA}
\affiliation{Department of Materials Science and Engineering, University of California, Berkeley, USA}

\author{Jian-Xin Zhu}
\affiliation{Theoretical Division and Center for Integrated Nanotechnologies, Los Alamos National Laboratory, Los Alamos, NM, USA}

\author{Eduardo H. da Silva Neto}
\affiliation{\mbox{Department of Physics, Yale University, New Haven, Connecticut 06520, USA}}

\affiliation{\mbox{Energy Sciences Institute, Yale University, West Haven, Connecticut 06516, USA}}

\author{Liang Wu}
\affiliation{Department of Physics and Astronomy, University of Pennsylvania, Philadelphia, PA 19104, USA}

\author{Jiun-Haw Chu}
\affiliation{Department of Physics, University of Washington, Seattle, Washington 98195, USA}

\author{Qimiao Si}
\affiliation{Department of Physics and Astronomy and Rice Center for Quantum Materials, Rice University, Houston, TX, 77005 USA}

\author{Ming Yi}
\email{mingyi@rice.edu}
\affiliation{Department of Physics and Astronomy and Rice Center for Quantum Materials, Rice University, Houston, TX, 77005 USA}

\date{\today}

\maketitle
\section{Abstract}
The ability to reversibly toggle between two distinct states in a non-volatile method is important for information storage applications. Such devices have been realized for phase-change materials, which utilizes local heating methods to toggle between a crystalline and an amorphous state with distinct electrical properties~\cite{Wuttig2007,Wang2021_phase,Wuttig2017,Kolobov2004,Poccia2011}. To expand such kind of switching between two topologically distinct phases requires non-volatile switching between two crystalline phases with distinct symmetries.
Here we report the observation of reversible and non-volatile switching between two stable and closely-related crystal structures with remarkably distinct electronic structures in the near room temperature van der Waals ferromagnet Fe$_{5-\delta}$GeTe$_2$. From a combination of characterization techniques we show that the switching is enabled by the ordering and disordering of an Fe site vacancy that results in distinct crystalline symmetries of the two phases that can be controlled by a thermal annealing and quenching method. Furthermore, from symmetry analysis as well as first principle calculations, we provide understanding of the key distinction in the observed electronic structures of the two phases: topological nodal lines compatible with the preserved global inversion symmetry in the site-disordered phase, and flat bands resulting from quantum destructive interference on a bipartite crystaline lattice formed by the presence of the site order as well as the lifting of the topological degeneracy due to the broken inversion symmetry in the site-ordered phase. Our work not only reveals a rich variety of quantum phases emergent in the metallic van der Waals ferromagnets due to the presence of site ordering, but also demonstrates the potential of these highly tunable two-dimensional magnets for memory and spintronics applications.

\clearpage

\section{Main}
Materials that can toggle between two states with distinct properties are important for information storage technology. Phase-change materials, for example, have been widely used for rewriteable optical data storage~\cite{Wuttig2007,Wang2021_phase,Wuttig2017,Jiang2018,Wang2017,Duan2021,Zhang2022_switch,Liu2023,Kolobov2004,Poccia2011,Fratini2010,Seyler2015,Eric2023,Liu2023_2,Du2021,Chiu2016,Tang2109,Zhang2019,Vergniory2019}. The key advantage is that the two phases are controlled by a non-volatile process, which is realized via a transient laser pulse that locally heats and changes the crystal structure, either resulting in a crystalline state or a quenched amorphous state. 2D van der Waals (vdW) materials is another class of material family whose properties are highly tunable, such as by electrostatic doping, optical illumination, or strain~\cite{2d_Burch2018,2d_Cheng2019,2d_Gong2017,2d_Huang2017,2d_Mak2019,Neto2009,Deng2020,Xu2020,Tokura2019, Bernevig2022,Chu2010,Yi2019,Mutch2019,Ricco2018,Cenker2022,Li2022,ZHENG2022,Ideue2019,Liang2017,Xi2013,Nikola2022}. They are valued not only for their versatile tunability but also the low dimensionality that allows exotic properties to arise due to quantum confinement. The advent of the concept of topology adds the potential to realize switching devices that go beyond resistive or optical readouts. As topology is often distinguished by crystalline symmetries, switching between two topologically distinct states can be realized via tuning knobs that change symmetries~\cite{Du2021}, most often achieved with a structural transition. However, such tuning knobs typically modulate temperature, electrostatic doping, strain, field, or pressure, all difficult to achieve in a non-volatile method. 

Here in this work, we demonstrate non-volatile reversible switching of two closely related crystal structural phases in the vdW ferromagnet Fe$_{5}$GeTe$_2$ via an annealing and quenching procedure. Fe$_{5}$GeTe$_2$ belongs to a class of Fe-based metallic vdW ferromagnets that exhibits relatively high Curie temperatures ($T_\mathrm{C}$ = 275 K to 330 K in the bulk limit)~\cite{Li2020_F5,Li2021_F5ARPES,Ly2021_F5ordering,Wu2021_F5arpes,Gao2020_F5domainwall,May2019_F5ACSNANO,May2019_F5PRM,Huang2022_F5ARPES,Zhang2020_F5,Zhang2022_F5dopedskymion,Junho2022_F345,Ribeiro2022_F5MBE,Chenxiang2022_F5doped}. 
Different from other widely studied 2D ferromagnets such as CrI$_3$ and Cr$_2$X$_2$Te$_6$ (X = Ge, Si)~\cite{2d_Huang2017,2d_Gong2017}, Fe$_{5}$GeTe$_2$ is air-stable and metallic, hence has been considered a top candidate for spintronics applications~\cite{2d_Huang2017,2d_Gong2017,2d_Cheng2019,May2019_F5PRM,May2019_F5ACSNANO}. 
The two phases share similar overall crystal structure but differ only in the ordering or disordering of an Fe vacancy site occupation that results in distinct crystalline symmetries. Second harmonic generation (SHG) measurements show the site-disordered phase to exhibit global inversion symmetry while the site-ordered phase breaks inversion symmetry, with intensity differing by a factor of 30. Remarkably, the electronic structures in the two phases are qualitatively distinct, as observed by angle-resolved photoemission spectroscopy (ARPES). From a combination of symmetry analysis and first principle calculations, we also provide a understanding of the key features of the observed electronic structure. In the site-disordered phase, we observe topological nodal lines that are compatible with the preserved global inversion symmetry, while in the site-ordered phase, we observe the lifting of the topological degeneracy due to the broken inversion symmetry as well as flat bands that are compatible with the quantum destructive interference of a bipartite crystalline lattice formed by the site-order. Our work not only demonstrates the exciting potential of using site order in the Fe-based 2D materials as a novel tuning knob to engineer and control correlated topological phases, but also reveals the potential of this class of 2D materials as a novel type of phase-change materials for non-volatile spintronics, memory or non-linear optical applications.


\subsection{Reversible switching of two distinct electronic structures}
\begin{figure}[]
\includegraphics[width=\textwidth]{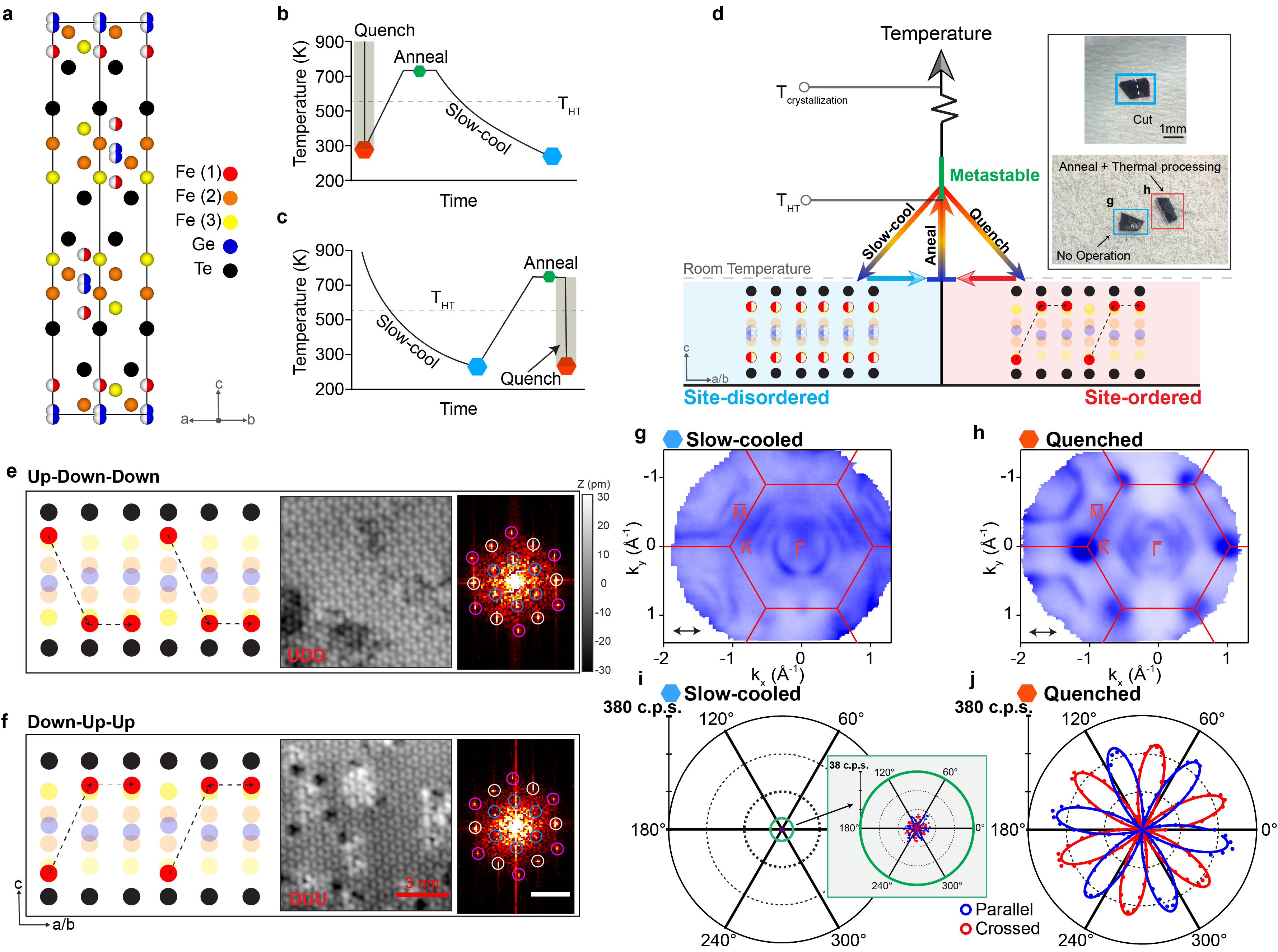} 
\caption{\textbf{Phase-tuning in Fe$_{5-\delta}$GeTe$_2$.}
\textbf{a}, Crystal structure of Fe$_{5}$GeTe$_2$ with atomic sites labeled. Fe(1) and Ge are modeled as split sites, marked by half-filled circles to represent 50\% occupancy. \textbf{b-d}, Schematic and procedures for tuning global symmetry via sublattice ordering. The two quantum phases can be switched by first annealing to above T$_{HT}$ =~550K and either quench or slowly cool to achieve the site-ordered or the site-disordered phase, respectively. The inset in \textbf{d} shows the real steps of tuning phases. \textbf{e-f}, Schematic model and STM topographic image of the $\sqrt{3}\times\sqrt{3}$ superstructure on Te termination. The right panels are Fourier transforms of the respective topographies with peaks corresponding to the lattice periodicity (white), a $\sqrt{3}\times\sqrt{3}$ superstructure periodicity (blue), and the second order of the $\sqrt{3}\times\sqrt{3}$ superstructure periodicity (pink). The inset scale bars are 15 nm$^{-1}$. The schematic lattice in \textbf{e} corresponds to a up-down-down (UDD) ordering of the Fe(1) atoms while that in \textbf{f} corresponds to a down-up-up (DUU) ordering of the Fe(1) sites. \textbf{g-h}, The Fermi surfaces of the two phases measured at 15 K using 114 eV LV photons, achieved by slow-cooling the crystal to room temperature from above T$_{HT}$ (g) and quenching from above T$_{HT}$ (h), respectively. \textbf{i-j}, Polarization-resolved SHG intensity on slow-cooled and quenched crystals measured at 5K. In both figures, the crossed and parallel configurations correspond to E(2\(\omega\))\(\bot\) E(\(\omega\)) and E(2\(\omega\)) \textbar\textbar{}E(\(\omega\)) respectively, while E(2\(\omega\)) and E(\(\omega\)) were simultaneously rotated in the crystal ab plane. c.p.s. stands for counts per second. The dots are experimental data and the solid curves are the fits by a six-fold sinusoidal function. Note that the radial axis for slow-cooled crystal in the inset is enhanced by a factor of 10.
}
\label{fig:fig1}
\end{figure}

Fe$_{5}$GeTe$_2$ belongs to a larger class of Fe-based metallic ferromagnets, Fe$_n$GeTe$_2$ (n=3 to 5)~\cite{Deng2018,Junho2022_F345,May2019_F5ACSNANO,May2019_F5PRM,wu2023spectral}, and is known to have a unique partially occupied split site~\cite{May2019_F5PRM,May2019_F5ACSNANO}. The crystal lattice of Fe$_{5}$GeTe$_2$ is rhombohedral (space group $R\overline{3}m$, No. 166)~\cite{May2019_F5PRM,May2019_F5ACSNANO}. The crystal structure consists of an ABC-stacking of the vdW slabs (Fig.~\ref{fig:fig1}a). Each slab consists of Fe and Ge sites sandwiched between layers of Te. In addition, each slab consists of three distinct Fe sites, marked as Fe(1), Fe(2), and Fe(3) in Fig.~\ref{fig:fig1}a. While Fe(2) and Fe(3) sites are fully occupied, Fe(1) sites are known to be split-sites where for each up-down pair within a single slab, they are either occupied in the up or down site~\cite{May2019_F5PRM,May2019_F5ACSNANO,Ly2021_F5ordering}. This choice of either the up or down site for Fe(1) pushes the Ge sites to also occupy a split site, where the site farther away from the occupied Fe(1) site is preferred. The choice of either occupying the up or down Fe(1) sites can be uncorrelated spatially or form an order depending on the rate at which the crystals are formed from growth~\cite{May2019_F5ACSNANO,May2019_F5PRM}. In particular, the occupancy of the Fe(1) sites could form an up-down-down (UDD) or down-up-up (DUU) pattern, resulting in a $\sqrt{3}\times \sqrt{3}$ superstructure~\cite{Ly2021_F5ordering}. This ordered occupancy is favored when the crystals are quenched from above a structural transition identified by previous literature as $T_\mathrm{HT}$ = 550 K while the random distribution is favored with slow cooling~\cite{May2019_F5PRM,May2019_F5ACSNANO}. For simplicity, we refer to the uncorrelated phase the site-disordered phase and the ordered phase the site-ordered phase. The ordering of the Fe(1) sites plays a crucial role in modifying the global symmetry of the crystal. In the site-disordered phase, the global inversion symmetry is preserved. This can be seen in Fig.~\ref{fig:fig1}a, where the inversion centers of each vdW slab is between the Ge split sites. In the site-ordered phase, the inversion symmetry is broken by the Fe(1) sites~\cite{Ly2021_F5ordering}. Such symmetry breaking has profound impact on the electronic structure, and as we will demonstrate, is the key to the tunability. 

To probe such an effect, we carried out ARPES measurements on crystals that were prepared in the two thermal methods. The measured Fermi surface (FS) of the slow-cooled crystals (Fig.~\ref{fig:fig1}g) and the quenched crystals (Fig.~\ref{fig:fig1}h) under the same measurement conditions are drastically different. In particular, the quenched crystals exhibit small pockets at the K points of the BZ, which are absent in the slow-cooled crystals. Instead, the slow-cooled crystals exhibit additional large pockets centered at the $\Gamma$ point. As we will show in detail in each of the two subsequent sections, the band dispersions leading to these FSs are significantly different, belonging to distinct topologically non-trivial phases. Before we discuss the electronic structure in depth, we first demonstrate the reversible non-volatile switching of these two phases. To confirm that it is the last thermal cooling step that dictates the electronic phase, we performed the following test (see Fig. S8 in the SI). First, we prepared the crystals by quenching them from above $T_\mathrm{HT}$ down to room temperature. Then we cut a crystal into halves and annealed a half piece to the metastable phase above $T_\mathrm{HT}$ and slowly cooled it back down to room temperature while leaving the other half untreated (Fig.~\ref{fig:fig1}b). The half pieces are then measured by ARPES. The electronic structure of the two halves are observed to be distinct, with the original quenched half identical to that shown in Fig.~\ref{fig:fig1}h, while the annealed and slow-cooled half identical to that presented in Fig.~\ref{fig:fig1}g. We have also checked the reverse process, which is to start with a crystal that was first formed via slow-cooling to room temperature, cut it in half, and annealing one half to above $T_\mathrm{HT}$ and then quenched in water (Fig.~\ref{fig:fig1}c). The subsequent ARPES measurement on the two halves again show the contrasting electronic structures, with the re-quenched half showing electronic structure identical to that in Fig.~\ref{fig:fig1}h and the original slow-cooled half identical to Fig.~\ref{fig:fig1}g. This procedure demonstrates that the key for the distinct electronic structures is the cooling rate in the final thermal treatment from above $T_\mathrm{HT}$. Hence we have demonstrated that there are two stable phases with drastically distinct electronic structures that can be reversibly switched in a non-volatile method. 

\subsection{Fe(1) site ordering as origin for distinct electronic phases}
As reported, there are three types of possible variations of the Fe$_{5-\delta}$GeTe$_2$ single crystals: Fe deficiency ($\delta$), stacking faults, and the formation of the \rtrt~Fe(1) site order~\cite{Ly2021_F5ordering,May2019_F5PRM,May2019_F5ACSNANO}. Since each pair of half crystals used above originate from the same original piece, the above procedure also rules out any difference in Fe deficiency as a potential cause for the difference in the electronic structure. Furthermore, we can also rule out the vdW stacking faults as a possible cause of the distinct electronic structure. From our transmission electron microscopy (TEM) images on the two types of crystals (see Fig. S1 in the SI), we do not observe any regular appearance of stacking faults in either the slow-cooled or quenched crystals. Both crystals exhibit ABC stacking, with occasional stacking faults between the vdW layers. Such rare occurrence cannot constitute a qualitative electronic structure distinction between the two types of crystals.

Therefore we are left with the appearance of the \rtrt~Fe(1) site order as the likely cause of the dichotomy of the electronic structure. First from single crystal x-ray diffraction (XRD) measurements, while the diffraction peaks corresponding to the $\sqrt{3}\times \sqrt{3}$ order are observed in the two types of crystals, their intensity relative to the Bragg peaks is reduced in the slow-cooled samples compared to those measured on a quenched crystal (see Extended Data Fig. S3). This suggests that while both site-disordered and site-ordered regions exist in the slow-cooled crystals, the population of the site-ordered regions is smaller.
To further confirm this, we carried out STM measurements on both quenched and slow-cooled crystals, revealing regions with a $\sqrt{3}\times \sqrt{3}$ superlattice with both UDD and DUU ordering of Fe(1) occupation sites (Fig. \ref{fig:fig1}e-f), consistent with previous STM reports on Fe$_{5-\delta}$GeTe$_2$~\cite{Ly2021_F5ordering}. While the field of view (on the order of 1 $\mu$m$^2$) of our STM measurements on quenched crystals showed only $\sqrt{3}\times \sqrt{3}$ ordered regions, similar measurements on slow-cooled crystals also showed disordered regions without the $\sqrt{3}\times \sqrt{3}$ superlattices. Interestingly, in slow-cooled crystals, these disordered regions dominate the field-of-view, surrounding small domains of $\sqrt{3}\times \sqrt{3}$ superlattices (see SI Extended Data Fig. S2), consistent with the XRD results. 
The existence of regions with $\sqrt{3}\times \sqrt{3}$ order in the two types of crystals revealed by STM is further confirmed by SHG measurement. We carried out polarization-dependent SHG measurements at 5 K on the two types of crystals. The quenched crystals reveal a 30 times stronger SHG signal compared to that of the slow-cooled crystals (Fig.~\ref{fig:fig1}i,j), note that to observe the tiny SHG signal (less than 10. c.p.s.) in the slow-cooled sample, a incident power of 4 mW with a 50 X objective is needed, which is just below the damage threshold, requiring a photon counter. As the SHG signal is contributed by the electric dipole (ED), \(I_{i}^{\text{ED}}(2\omega) \propto \left|{\Sigma}_{jk}\chi_{ijk}^{\text{ED}}E_{j}(\omega)E_{k}(\omega) \right|^{2}\), where \(k\) and \(\omega\) are the wavevector and frequency of incident beam respectively, \(\chi\) is the nonlinear susceptibility tensor and \(i,\ j,k,l\) are Cartesian coordinate indices, it is a sensitive probe of the presence of inversion-symmetry-breaking. On one hand, for the quenched crystals, the clear presence of inversion symmetry breaking is consistent with the formation of the \rtrt~order. On the other hand, in the slow-cooled crystals dominated by regions with random Fe(1) site occupancy, the electric dipole contribution to SHG would be forbidden due to the preserved global inversion symmetry while only a smaller electric quadrupole (EQ) SHG contribution following the three-fold rotational symmetry would be allowed, \(I_{i}^{\text{EQ}}(2\omega)\propto\left| {\Sigma}_{jkl}\chi_{ijkl}^{\text{EQ}}k_{j}E_{k}(\omega)E_{l}(\omega) \right|^{2}\) under normal incidence. The much enhanced SHG signal in the quenched crystals is consistent with both the STM and XRD observations. Hence, we associate the electronic structure measured on the quenched crystals to that of the Fe(1) site-ordered phase and that measured on the slow-cooled crystals to that of the Fe(1) site-disordered phase. As we will demonstrate subsequently in the discussion section, the crystal symmetries for the site-disordered and site-ordered phases are highly compatible with the topological band dispersions that we observe from ARPES.

\subsection{Nodal lines in the site-disordered phase}
\begin{figure}[]
\includegraphics[width=\textwidth]{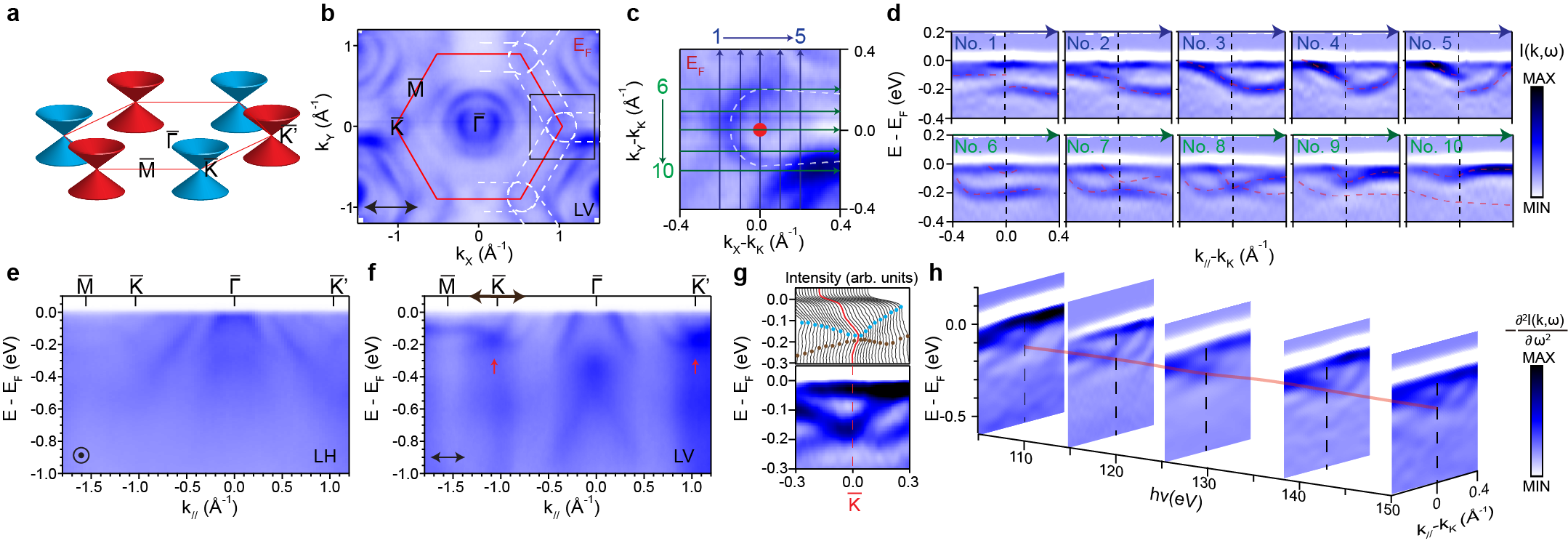} 
\caption{\textbf{Topological nodal lines in the slow-cooled site-disordered phase.}
\textbf{a}, Schematic showing Dirac cones at K and K' on a 2D projected BZ plane. \textbf{b}, Fermi surface mapping with LV polarized photons. The red line outlines the 2D BZ boundary and the white dashed line marks the electron pocket surrounding the BZ boundary. \textbf{c}, Zoomed in view of the box in \textbf{b}. \textbf{d}, Dispersions measured across the Dirac node, showing the crossing bands and the opening of a gap away from the Dirac node. \textbf{e-f}, Dispersions measured along the $\bar{M}-\bar{K}-\bar{\Gamma}-\bar{K'}$ direction with LH and LV polarization, respectively. The red arrows in \textbf{f} point to the Dirac crossings. \textbf{g}, Energy distribution curves (EDCs) and the second energy derivative cut within the momentum range marked by black arrow in \textbf{f}. \textbf{h} Photon energy dependence of the cut near the $\bar{K}$ point. The red solid line shows the Dirac nodal line along \kz. All data from \textbf{b} to \textbf{g} were taken with 132 eV photons.
}
\label{fig:fig2}
\end{figure}

Next, we present in detail the key features in the measured electronic structure of the site-disordered phase achieved from slow-cooling the crystals. The FS in the ferromagnetically ordered state is shown in Fig.~\ref{fig:fig2}b, consisting of several circular Fermi pockets centered at the BZ center and elliptical pockets surrounding the $\bar{K}$- $\bar{M}$- $\bar{K'}$ BZ boundaries, as highlighted by white dashed lines. To understand these features, we measured the electronic dispersions along the high symmetry direction of $\bar{M}-\bar{K}-\bar{\Gamma}-\bar{K}$ (Fig.~\ref{fig:fig2}e-f). First, we observe a number of hole bands centered at the $\bar{\Gamma}$ point, giving rise to the observed circular Fermi pockets. Interestingly, near the $\bar{K}$ point, we also observe a band crossing near -0.18~eV. The crossing can be better visualized from the energy distribution curves (EDC) as well as second energy derivatives of the raw spectra (Fig.~\ref{fig:fig2}g).
The nature of the crossing can be further demonstrated from a series of cuts of the measured band dispersions close to the $\bar{K}$ point. In Fig.~\ref{fig:fig2}d, both horizontal cuts (cuts 1 to 5) and vertical cuts (cuts 6 to 10) in the crossing region reveal two bands that cross near the $\bar{K}$ point and become gapped away from $\bar{K}$. 

Having demonstrated the band crossing at the $\bar{K}$ points in the in-plane direction, we also examine the dispersion along the out-of-plane direction (k$_z$) by varying the photon energy. As the inter-layer interactions in vdW materials are quite weak, we do not observe strong variation along k$_z$ (see SI Extended Data Fig. S6). For a range of photon energies that probes a range much beyond that of a single BZ along k$_z$, we always observe the crossing near the $\bar{K}$ point (Fig.~\ref{fig:fig2}h), hence the in-plane nodal crossing takes the form of nodal lines along the out-of-plane direction. Taking these findings together, our ARPES data reveal the existence of nodal lines along the BZ boundaries. As these nodal lines are observed in a ferromagnetic phase, the time-reversal symmetry is broken and hence the spin degree of freedom is quenched, giving rise to two-fold degenerate lines.

\subsection{Flat bands in the site-ordered phase}

\begin{figure}[]
\includegraphics[width=0.5\textwidth]{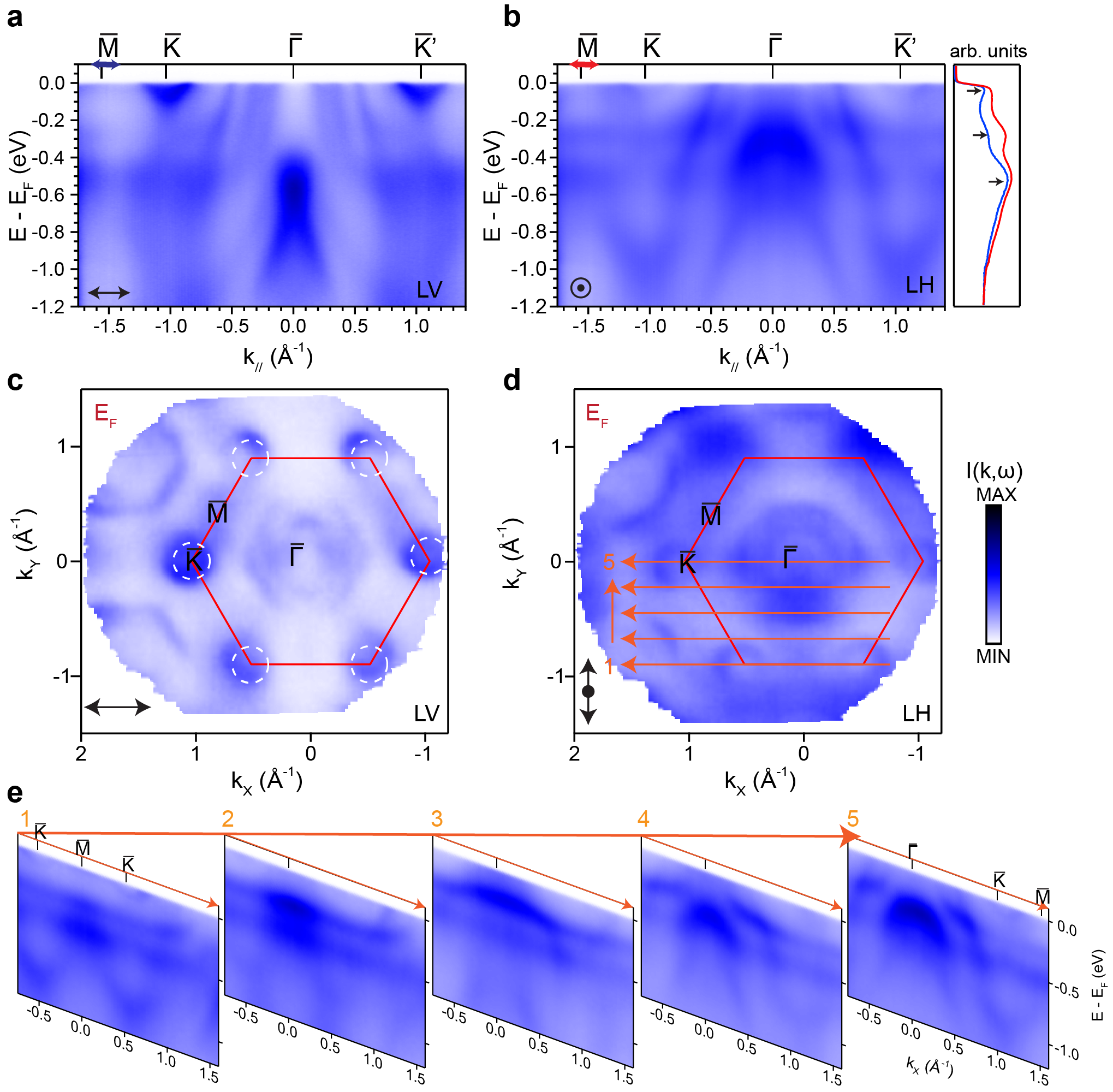} 
\caption{\textbf{Flat bands in the quenched site-ordered phase.}
\textbf{a,b}, 
measured energy-momentum dispersions along the $\bar{M}-\bar{K}-\bar{\Gamma}-\bar{K'}$ direction with LV and LH polarized photons. The EDCs are integrated from a small range around $\bar{M}$ marked by the arrows in \textbf{Fig. 3a, b}, showing three peaks that correspond to the location of the flat bands. \textbf{c,d}, Corresponding Fermi surfaces measured with LV and LH polarizations, respectively. Data in \textbf{c} were measured under the same geometry as \textbf{Fig. 2b}. \textbf{e}, Measured dispersions along a series of parallel cuts within the first BZ as shown in \textbf{d}. All data from \textbf{a} to \textbf{e} were taken with 114 eV photons.
}
\label{fig:fig3}
\end{figure}

Having presented the existence of the nodal lines in the site-disordered phase, we now focus on the observed electronic structure of the site-ordered phase. Figure~\ref{fig:fig3} summarizes the measured electronic structure of quenched crystals. 
Instead of the elliptical Fermi pockets surrounding the $\bar{K}-\bar{M}$ direction resulting from the Dirac crossing at $\bar{K}$ points in the slow-cooled crystals, the quenched crystals exhibit circular pockets at the $\bar{K}$ points. 
This distinction can be further seen from dispersions measured along the high symmetry direction $\bar{M}-\bar{K}-\bar{\Gamma}-\bar{K}$. In stark contrast to that measured for the site-disordered phase (Fig.~\ref{fig:fig2}), the site-ordered crystals show electron bands at the $\bar{K}$ points with clear band bottoms and no band crossings, and hence the absence of the nodal lines observed in the site-disordered crystals.

More interestingly, three flat bands are observed in the site-ordered crystals that are not observed in the site-disordered crystals. We first illustrate them along the $\bar{K}-\bar{\Gamma}-\bar{K}-\bar{M}$  direction, captured in measurements under both linear horizontal (LH) and linear vertical (LV) polarizations (Fig.~\ref{fig:fig3}a-b). The location of the flat bands can be identified as peaks in the integrated EDCs from both polarizations, at \ef, -0.2~eV, and -0.6~eV. Beyond the high symmetry direction, the flat bands are observed to persist across a large region of the BZ. We illustrate this from five cuts measured across the in-plane BZ (Fig.~\ref{fig:fig3}d-e). The flat band near \ef~could be clearly seen along the $\bar{K}-\bar{M}- \bar{K}$ direction as shown on cut 1. When the $\bar{\Gamma}$ point is approached from cut2 to cut5, the flat band near \ef~shifts to above \ef~and could no longer be observed. The second and third flat bands located at -0.2~eV and -0.6~eV are flat throughout the BZ except where they hybridize with the dispersive bands near the $\bar{\Gamma}$ point. We note that this hybridization indicates that these flat dispersions are intrinsic to the crystal and cannot be due to disorders or impurities that would otherwise form momentum-independent states that do not interact with intrinsic band structure. Furthermore, we carried out photon energy-dependent measurements, where the flat bands are observed to persist across \kz~(see SI), consistent with the 2D nature of the vdW materials.


\subsection{Topology for the distinct electronic phases}
\begin{figure}[]
\includegraphics[width=0.65\textwidth]{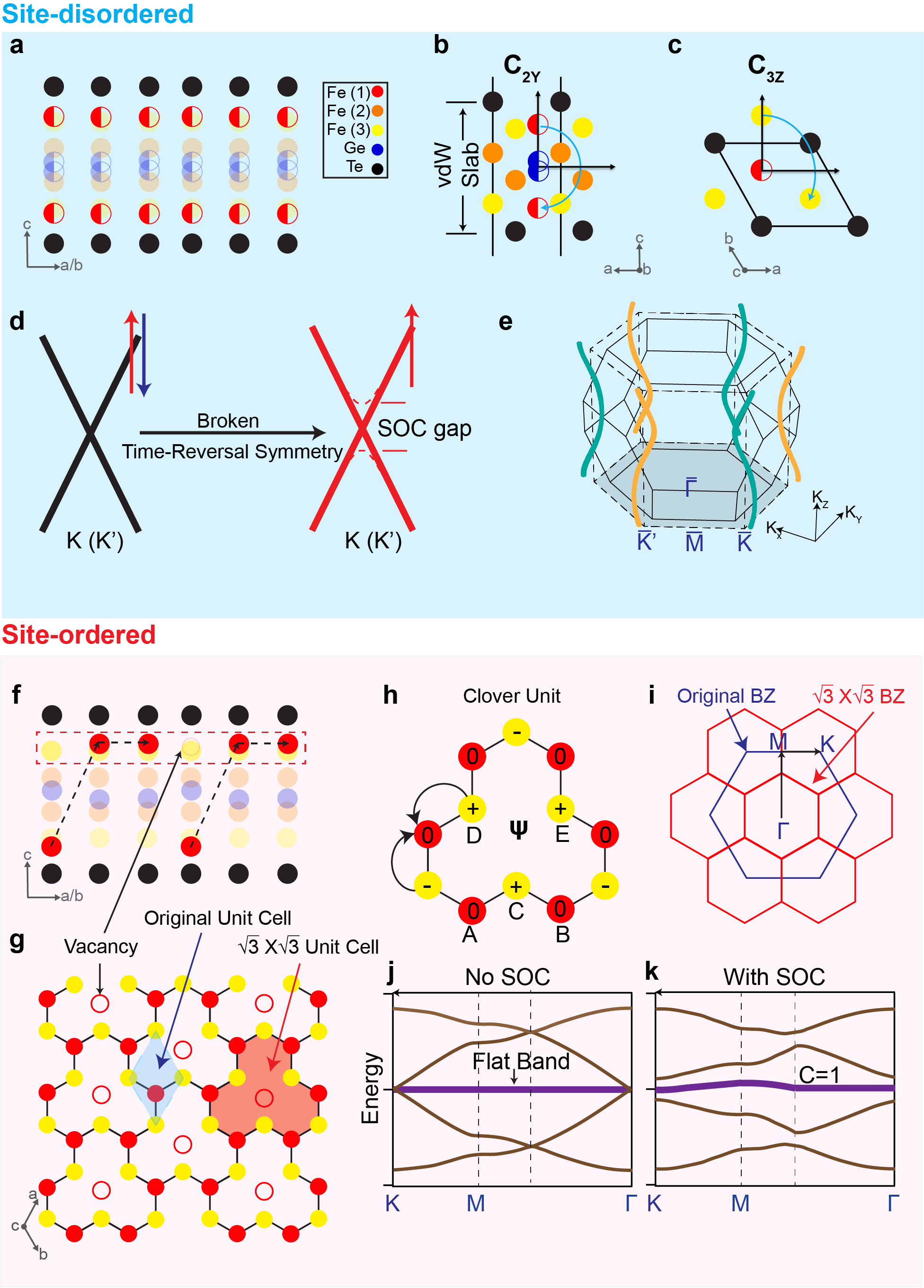} 
\caption{\textbf{Effect of the Fe(1) site ordering on inversion-symmetry and quantum destructive interference.}
\textbf{a}, Illustration of one vdW slab of the Fe(1) site-disordered phase, in which the Fe(1) sites are randomly distributed such that global inversion symmetry is preserved. \textbf{b-c}, The schematic view of the lattice showing the C$_{2y}$ and C$_{3z}$ symmetries in the Fe(1) site-disordered phase, respectively. \textbf{d}, The tight-binding model showing Dirac crossings at K (K') protected by the above symmetries and the associated two-fold topological crossing due to the ferromagnetic order and the gap by SOC. \textbf{e}, The helical topological nodal lines induced by the ABC stacking of the vdW slabs that results in winding of the nodal lines with opposite chiralities around $\bar{K}$ and $\bar{K'}$. \textbf{f}, Illustration of one vdW slab of the Fe(1) ordered-site phase, in which the Fe(1) is DUU-ordered, forming a bipartite crystalline lattice. \textbf{g}, The Fe(1) and the nearest neighbor Fe(3) sites form a clover lattice as shown in the red dashed box in \textbf{f}. \textbf{h}, Single clover unit showing the destructive interference of the hopping amplitude at the Fe(1) and Fe(3) sites due to the alternating sign of the Wannier phase, leading to a localization of the electronic wavefunction. The sites and wavefunction amplitudes are labeled on the corresponding atoms. \textbf{i}, The original (blue) and the $\sqrt{3}\times\sqrt{3}$ superstructure (red) BZ. \textbf{j-k}, Tight-binding model for the clover lattice without and with spin orbital coupling (SOC), respectively. With SOC, the flat band is gapped with a Chern number of 1, and is hence topologically nontrivial.
}
\label{fig:fig4}
\end{figure}

The drastically distinct electronic structures of the two types of crystals, with one exhibiting two-fold nodal lines and the other flat dispersions, belong to distinct topological states. Here we show that they can be understood from the symmetries dictated by the site-disordered or site-ordered phases, respectively. We first discuss the case of the site-disordered phase where we observe nodal lines at the K points. For a single vdW slab with 50\% occupation of the Fe(1) sites, the crystalline symmetry belongs to the centrosymmetric space group $P\bar{3}1m$ (No. 164). Here, the crystal has both two-fold rotational symmetry about the $y$ axis ($C_{2y}$) (Fig.~\ref{fig:fig4}b) and three-fold rotational symmetry about the $z$ axis ($C_{3z}$) (Fig.~\ref{fig:fig4}c), similar to the case of graphene. The momentum point $K$ ($K'$) is invariant under these two symmetry operations and allows the existence of a 2D irreducible representation. In the ferromagnetic phase where time-reversal symmetry is broken, the spin-polarized bands in the ferromagnetic state can be regarded as spinless states and would cross at the K~(K') points, where the two-fold degeneracy comes from the orbital degree of freedom (see SI for a discussion of the orbitals), leading to a symmetry-enforced ferromagnetic Dirac crossing. To demonstrate this, we built an effective tight binding model considering the different Fe 3d orbitals and show that such a crossing is indeed protected at the K~(K') point (see SI for a full discussion of the tight-binding model). 
When we incorporate spin-orbit coupling (SOC), (see SI), in general, the SOC can be expressed as $H_{SO}=\lambda_{SO} {\bm L}\cdot {\bm S}$, where $\bm{L}$ and $\bm{S}$ are the angular and spin momenta, respectively. In a ferromagnetic system, ${\bm S} \sim \langle {\bm S} \rangle$ plays the role of an effective Zeeman splitting field in the orbital basis. Since the direction of the magnetic moment is along the $z$-direction~\cite{May2019_F5PRM,May2019_F5ACSNANO}, which is parallel to the direction of the orbital angular momentum, the SOC would lift the two-fold degeneracy at K and K'. The band structure calculation from the tight binding model with SOC is illustrated in Fig.~\ref{fig:fig4}d. This is consistent with our experimental observation of the crossing at K points except that the gap due to SOC is not resolved in the experiment due to the energy resolution. The appearance of this degeneracy at K points is similar to that reported in the related Fe$_3$GeTe$_2$, where the topological nodal lines are theoretically identified to give rise to a large anomalous Hall effect~\cite{Kim2018}, but difficult to resolve in the ARPES measured dispersions. Here in Fe$_5$GeTe$_2$, they are clearly observed.

Having understood the single layer case, we now consider the bulk system of the site-disordered phase. In a simple hypothetical AAA stacking scenario, the hopping along the z direction extends the original 2D hexagonal BZ into a 3D hexagonal prism and would extend the topological crossings at K and K' to nodal lines along the K-H direction. This is protected by a combination of $C_{3z}$ and $PT$ symmetries. For the real ABC stacking of the layers, the BZ changes from a hexagonal prism into the BZ of a rhombohedral space group (Fig.~\ref{fig:fig4}e). Due to the ABC stacking of the layers, the K-H direction is no longer a high symmetry line of the BZ. Instead, the topological crossings at each $k_z$ plane shift away from the K and K' points, forming helical nodal lines that wind around K-H, 
where the magnitude of the shift is proportional to the strength of the interlayer hopping, similar to helical nodal lines reported in other ABC-stacked materials including Fe$_3$Sn$_2$~\cite{Heikkila2011,Ye2018,Fang2022}. Here in Fe$_5$GeTe$_2$, due to the weak vdW interlayer coupling, the in-plane deviation of the crossing from the K-H line is too small to be experimentally resolved. Hence we cannot directly observe the winding but only observe nodal lines near K-H. 

In addition to tight-binding calculations, we also carried out density functional theory (DFT) calculations to check for the symmetry-enforced crossings (see SI Extended Data Fig. S4a-c). To demonstrate the importance of the globally preserved inversion symmetry of the random Fe(1) occupation, we carried out the following comparison. First, we calculated the band structure for the Fe(1) sites all occupying the up sites (UUU). The inverted case is the structure with Fe(1) sites all occupying the down sites (DDD). The average of the two from directly overlapping the UUU and DDD band structures would give an average stoichiometry of Fe$_5$GeTe$_2$. We note that while such structure does not exist in the crystal, it mimics the site-disordered phase except it lacks inversion symmetry. To directly compare this calculation with an inversion symmetric structure, we also calculated the band structure of a crystal structure with both up and down sites fully occupied, giving a stoichiometry of Fe$_6$GeTe$_2$. By comparing calculations without and with SOC, only the inversion symmetric Fe$_6$GeTe$_2$ shows band crossings at the K point that open up a gap with the inclusion of SOC, demonstrating the symmetry-enforced nature of the topological nodal lines. The UUU and DDD band structures do not exhibit such kind of band crossing, confirming that the presence of global inversion symmetry is consistent and also required for the observed topological nodal lines.  


Finally, we discuss the site-ordered phase. Consistent with the inversion symmetry breaking observed by SHG, we no longer observe the topological crossings at the K point. Such inversion symmetry breaking is consistent with the \rtrt~order caused by the Fe(1) site ordering. Interestingly, for such DUU occupation order of the Fe(1) site (Fig.~\ref{fig:fig4}f), the shortest bond occurs between the Fe(1) sublayer and the adjacent Fe(3) sublayer~\cite{May2019_F5PRM,May2019_F5ACSNANO}. The in-plane projection of these two sublayers form a clover unit pattern, with the center being the missing Fe(1) site, as shown in Fig.~\ref{fig:fig4}g-h. 
Considering only the nearest neighbor hopping, $t_1$, which is between the red Fe(1) sites and yellow Fe(3) sites, the lattice is manifested as a bipartite crystalline lattice (BCL), in which the lattice is categorized into two sublattices with different numbers of atoms (Fig.~\ref{fig:fig4}h). BCLs are predicted to be a generic platform to realize destructive interference of the electronic wavefunction and further lead to flat bands~\cite{Regnault2022,Calugaru2022}, but have never been directly observed in bulk materials.

To see this clearly, we consider the Hamiltonian for the single orbital clover lattice with nearest neighbor hopping
\begin{equation}
    H(\kbf) = \begin{pmatrix}
    \mathbf{0_{2\times 2}} & \mathcal{H}_{\kbf} \\
    \mathcal{H}_{\kbf} & \mathbf{0_{3\times 3}}
    \end{pmatrix},
\end{equation}

where $\mathcal{H}_{\kbf}$ is the hopping matrix between two sublattices (see explicit form in Eq.~\ref{eq:fb_ham} in methods). Given that $\mathcal{H}_{\kbf}$ is a $3\times2$ rectangular matrix, the Hamiltonian contains at least $3-2=1$
zero modes for all $\kbf$. The band structure is shown in Fig.~\ref{fig:fig4}j, after diagonalizing the Hamiltonian. The correspinding localized wavefunction for the flat band is:
\begin{equation}
    \psi(k_x, k_y) = \begin{pmatrix}
    0 & 0 & e^{\frac{k_y}{3}} - e^{-\frac{2}{3}k_y} & e^{-\frac{k_x}{2\sqrt{3}} - \frac{k_y}{6}} -  e^{\frac{k_x}{\sqrt{3}} + \frac{k_y}{3}} & e^{\frac{k_x}{2\sqrt{3}} - \frac{k_y}{6}} - e^{-\frac{k_x}{\sqrt{3}} + \frac{k_y}{3}}
    \end{pmatrix},
\end{equation}
leading to a real space Wannier function as shown in Fig.~\ref{fig:fig4}h. The Wannier amplitude is identically zero on the red sites because of the destructive interference effect.
When SOC is incorporated, the flat band gains dispersion and also acquires a finite Chern number, and becomes topologically non-trivial. The consideration for different orbital groups is also provided in the methods. Such kind of destructive-interference induced flat bands have been discussed in kagome~\cite{Regnault2022,Calugaru2022,Yin2018,Kang2020,Kang2020_2} and pyrochlore~\cite{Regnault2022,Calugaru2022} lattices. Here in quenched Fe$_5$GeTe$_2$, they are directly the result of the geometrically frustrated lattice formed by the Fe(1) occupation site ordering enabled by the quenching process. 

We carried out DFT calculations to check for the BCL flat bands associated with the clover lattice (see SI Extended Data Fig. S4d). To mimic the site-disordered phase, we overlap the band structures for the UUU and DDD structures and compare it to the band structure calculated with the site-ordered phase with the \rtrt~order. A direct comparison of the two shows that no flat bands are observed in the UUU+DDD calculation but flat bands are observed for the site-ordered phase. We can also unfold the band structure of the \rtrt~order back to the original unfolded BZ to compare more directly with the observed dispersions, and find reasonable agreement (see SI Extended Data Fig. S5). Projection of the density of states unto the different Fe sites also show that the peaks corresponding to the flat bands have large contributions from the Fe(1) and Fe(3) sites that form the clover lattice. This demonstrates that the flat dispersions that we observe only in the site-ordered phase and not in the site-disordered phase are associated with the clover unit that only forms with the Fe(1) site order, and is a direct result of the quantum destructive interference of the bipartite crystalline lattice.

\subsection{Discussion}
Taking all experimental and theoretical evidence presented together, we have demonstrated the reversible switching of two remarkably distinct electronic structures ascribed to two closely-related crystalline phases via a non-volatile thermal process in Fe$_5$GeTe$_2$. The capability is enabled by the Fe(1) site ordering that changes the crystal symmetries leading to distinct topological characteristics. On one hand, the random occupation of the Fe(1) sites leads to global inversion symmetry that allows symmetry-enforced topological nodal lines, which are observed to be lifted when the inversion-symmetry breaking order forms. On the other hand, the formation of the site order creates a bipartite crystalline lattice that localizes electronic states to form flat bands, which are observed to be destroyed with the breaking of the site order. Our findings indicate that the Fe$_5$GeTe$_2$ system is a rich system for probing and understanding topology in the correlated regime. As Fe$_5$GeTe$_2$ is known to exhibit high Curie temperature, it would be interesting to compare the magnetic properties of the two phases, including manipulation of the topological nodal lines in the site-disordered phase and the role of the topological  flat bands for magnetism in the site-ordered phase. Fe$_5$GeTe$_2$ is also an interesting system to probe from the order-disorder perspective. As our STM results show that the slow-cooled samples exhibit domains of ordered regions, it would be interesting to understand how the domains form and propagate in the cooling process as a function of cooling rate, especially given that Fe$_5$GeTe$_2$ behaves counterintuitively in that the ordered phase is preferred via quenching. Such studies would benefit from the vast expertise developed for probing and understanding order-disorder formation in other quantum materials~\cite{Wuttig2007,Wang2021_phase,Wuttig2017,Jiang2018,Wang2017,Duan2021,Zhang2022_switch,Liu2023,Kolobov2004,Poccia2011,Fratini2010,Seyler2015,Eric2023,Liu2023_2}. Finally, the non-volatile switch our work exemplifies promises versatile settings to apply a novel design principle, viz to utilize the cooperation of crystalline symmetry and strong correlations to produce new correlated topological materials~\cite{ChenLei2022}.

Aside from fundamental physics, our work also indicates that Fe$_5$GeTe$_2$ has great potential for applications. Skyrmions have recently been reported in this class of Fe-based vdW ferromagnets~\cite{F3_sky_Birch2022,F3_sky_Ding2020,F3_sky_Yang2020,F5_sky_Brian2023,F5_sky_Fujita2022,Schmitt2022_F5skyrmion}. As skyrmions are stablized by Dzyaloshinsky–Moriya interaction, which is only allowed when inversion symmetry is broken, there has been debates on how to understand the appearance of skyrmions in these seemingly centrosymmetric crystals. In the case of Fe$_3$GeTe$_2$, this has been explained via random Fe deficiencies that on average occur asymmetrically in the crystal~\cite{Anirban2022}. In the case of (Fe,Co)$_5$GeTe$_2$, this is ascribed to AA' stacking of the vdW layers~\cite{meisenheimer2022controlled}. Here we show that the two phases have clean distinction on inversion symmetry via the site-ordering process, hence provides a platform to potentially control skyrmion formation. The process by which we demonstrate the switching--heating and cooling all above room temperature--is similar to that already commercially used for phase-change materials such as Ge$_2$Sb$_2$Te$_5$~\cite{Wuttig2007,Wuttig2017,Kolobov2004}. Different from phase-change materials, we only need to surpass a submelting temperature where the Fe(1) sites are mobilized instead of having to achieve the melting and crystallization temperature. Techniques such as local laser heating can be explored for spatial writing of the two phases especially given that the overall crystal structures are compatible. This is in contrast to some vacancy ordered materials such as K$_x$Fe$_{2-y}$Se$_2$ where the metallic regions are structurally unstable and only appears as microstructures amidst the insulating vacancy ordered phases~\cite{Bao_2013,Ding2013}. The heating and quenching process that we utilize is non-volatile and above room temperature, which is advantageous compared from those controls that require the presence of field, strain, pressure, or current. Nevertheless, modifying the Fe(1) sites and their vacancies appears to have lower energy barrier than re-crystallization, suggesting that electrical current, photo illumination or other commonly utilized switching methodologies could also be explored for this 2D vdW material.

Finally, the concept of using vacancy order-disorder to realize distinct topological phases goes beyond Fe$_5$GeTe$_2$. Phase change via order-disorder has been explored extensively for realizing switches based on electrical or optical properties. Here we demonstrate the concept that vacancy order can be utilized to change the crystalline symmetries of two otherwise energetically similar ground states with dramatically distinct consequences on their topological character. A large base of quantum materials are known to exhibit vacancies or site disorder. The consideration of the symmetries of these phases may open up new routes towards realizing exotic topological phases as well as novel spintronics applications.
\section{Data Availability}

The data that support the findings of this study are available from the corresponding author upon reasonable request.


\section{Acknowledgments}
The authors acknowledge insightful discussions with Kai Sun, Luis Balicas, and Alex Frano. This research used resources of the Advanced Light Source, the Stanford Synchrotron Radiation Lightsource, and the National Synchrotron Light Source-II, all U.S. Department Of Energy (DOE) Office of Science User Facilities under contract Nos. DE-AC02-05CH11231, AC02-76SF00515 and No. DE-SC0012704, respectively. Rice ARPES work is supported by the U.S. DOE grant No. DE-SC0021421 and the Gordon and Betty Moore Foundation’s EPiQS Initiative through grant no. GBMF9470. The theory work at Rice is primarily supported by the U.S. DOE, BES, under Award No. DE-SC0018197 (L.C., symmetry analysis), by the AFOSR under Grant No. FA9550-21-1-0356 (C.S., electronic structure construction), and by the Robert A. Welch Foundation Grant No. C-1411 (Q.S.).
Work at Los Alamos was carried out under the auspices of the U.S. Department of Energy (DOE) National Nuclear Security Administration (NNSA) under Contract No. 89233218CNA000001, and was supported by LANL LDRD Program, UC Laboratory Fees Research Program (Grant Number: FR-20-653926), and in part by the Center for Integrated Nanotechnologies, a DOE BES user facility. The development of the SHG photon counter is supported by the Army Research Office and was accomplished under grant no. W911NF-19-1-0342. The sample exfoliation is based upon work supported by the Air Force Office of Scientific Research under award number FA9550-22-1-0449. Q.D. is supported by the NSF EPM program under grant no. DMR-2213891. L.W. acknowledges the support by the Air Force Office of Scientific Research under award no. FA9550-22-1-0410. TEM study is supported by Welch Foundation (C-2065-20210327). The authors acknowledge the use of the Electron Microscopy Center at Rice. The work at LBL and UC Berkeley was funded by the U.S. Department of Energy, Office of Science, Office of Basic Energy Sciences, Materials Sciences and Engineering Division under Contract No. DE-AC02-05-CH11231 (Quantum Materials program KC2202). Research conducted at the Center for High-Energy X-ray Sciences (CHEXS) is supported by the National Science Foundation (BIO, ENG and MPS Directorates) under award DMR-1829070. Materials synthesis at UW was supported as part of Programmable Quantum Materials, an Energy Frontier Research Center funded by the U.S. Department of Energy (DOE), Office of Science, Basic Energy Sciences (BES), under award DE-SC0019443

\section{Author Contributions}
The project was initiated and organized by M.Y. The single crystals were grown by P.M., Y.S., X.C., Y.H, C.H, X.X and J.C. The ARPES measurements and analyses were carried out by H.W., J.W.H., J.S.O., R.J.B. and M.Y. with the help of D.H.L., M.H., S.-K.M., A.F., J.D., T.Y. and E.V. The tight binding model and symmetry analyses were proposed and carried out by L.C., C.S. and Q.S. The first principle calculations were carried out by B.G.J. and J.Z. The SHG were carried out by Q.D. and L.W. The STM measurements were measured by K.S. and E.dsN. The x-ray diffraction was done by J.R. The TEM were measured by C.S. and Y.-M.H. The sample annealing and quenching process and characterization were carried out by P.M., J.C., X.C., Y.X., B.G., X.T., M.K., H.W. and P.D. The manuscript was written by H.W. and M.Y. and contributed by all the authors.

\section{Competing Interests}

The authors declare no competing interests.

\bibliographystyle{naturemag}
\bibliography{bib}

\newpage

\end{document}